\documentclass{cpcauth}
\usepackage{graphicx}
\usepackage{url}
\newcommand{\EQ}{\begin{equation}}
\newcommand{\EN}{\end{equation}}
\newcommand{\EQA}{\begin{eqnarray}}
\newcommand{\ENA}{\end{eqnarray}}

\newcommand{\Eq}[1]{Eq.~(\ref{#1})}

\newcommand{\Fig}[1]{Fig.~\ref{#1}}

\newcommand{\Tab}[1]{Table~\ref{#1}}

\newcommand{\bra}[1]{\langle #1\rangle}

\newcommand{\meanBB}{\overline{\bf{B}}}
\newcommand{\meanJJ}{\overline{\bf{J}}}
{}
{}
{}
{}
{}
{}
{}
\newcommand{\uu}{{\bf{u}}}
\newcommand{\BB}{{\bf{B}}}
\newcommand{\JJ}{{\bf{J}}}
\newcommand{\jj}{{\bf{j}}}
\newcommand{\AAA}{{\bf{A}}}

\newcommand{\bb}{{\bf{b}}}

\newcommand{\EE}{{\bf{E}}}

\newcommand{\nab}{\mbox{\boldmath $\nabla$} {}}

\newcommand{\dd}{{\rm d} {}}

\def\half{{\textstyle{1\over2}}}

\begin{document}

\title{Hydromagnetic turbulence in computer simulations}
\author{A. Brandenburg$^{1,2}$ and W. Dobler$^{2,3}$}
$^1$NORDITA, Blegdamsvej 17, DK-2100 Copenhagen \O, Denmark\\
$^2$Department of Mathematics, University of Newcastle, NE1 7RU, UK\\
$^3$Kiepenheuer-Institut f\"ur Sonnenphysik, Sch\"oneckstr. 6, 79104 Freiburg, Germany

\maketitle


\section{Introduction}

Hydromagnetic processes play an important role in many
astrophysical systems (e.g.~stars, galaxies, accretion discs).
This is because the medium is hot enough to be partially
or fully ionized. Because of the huge scales involved the
medium is usually turbulent, provided there is an instability
(shear, convection) facilitating
the cascading of energy down to small scales.

In turbulence research it has been a long standing tradition to solve
the hydrodynamic equations using spectral schemes which have the lowest
possible discretization error. Spectral schemes are particularly
useful for incompressible problems where one needs to solve a
Poisson-type equation for the pressure. However, spectral schemes
are no longer optimal in many astrophysical circumstances where flows
are generally compressible. Lower order spatial derivative schemes
are generally unacceptable in view of their low overall accuracy, even
when schemes are used where mass, momentum, and energy are conserved
to machine accuracy. On massively parallel machines, on the other hand,
spectral schemes are difficult to make run efficiently. High order finite
difference schemes are therefore a useful compromise. Such schemes
can yield almost spectral-like accuracy.

Our code
uses centered finite differences which make the adaptation to other
problems simple. Since the code is not written in conservative form,
conservation of mass, energy and momentum can be used to monitor the
quality of the solution. A third order Runge-Kutta scheme with $2N$
storage \cite{2Nstorage} is used for calculating the time advance.

\section{Advantages of high-order schemes}

Spectral methods are commonly used in almost all studies of ordinary
(usually incompressible) turbulence. The use of this method is justified
mainly by the high numerical accuracy of spectral schemes. Alternatively,
one may use high-order finite differences that are faster to compute
and that can possess almost spectral accuracy.
In astrophysics, high-order compact finite differences \cite{Lele92}
have been used to model stellar convection \cite{NS90,Ref-3} and shear
flows in accretion discs \cite{BNST95}.
In contrast to explicit finite differences, compact finite
differences \cite{Lele92} have a smaller coefficient in the
leading error, even if both schemes are of the same order.
However, compact schemes are still {\it nonlocal} in the sense that each
point affects every other point, which enhances communication. This is
the main reason why we adopt {\it explicit} centered finite differences.

In this section we demonstrate, using simple test problems, some of
the advantages of high-order schemes.
The explicit formulae for first and second derivatives are
\EQA
f'_i=(-f_{i-3}+9f_{i-2}-45f_{i-1}
\nonumber \\
+45f_{i+1}-9f_{i+2}+f_{i+3})/(60\delta x),
\ENA
\EQA
f''_i=(2f_{i-3}-27f_{i-2}+270f_{i-1}-490f_i
\nonumber \\
+270f_{i+1}-27f_{i+2}+2f_{i+3})/(180\delta x^2).
\ENA
Full details of these schemes, including formulae for the
boundaries, can be found in Ref.~\cite{Ref-1}.
This scheme was also used in recent applications 
to the problem of resistively limited growth in
models of stellar dynamos; see Ref.~\cite{Ref-4}.

\begin{figure}[t!]\centering\includegraphics[width=0.99\columnwidth]{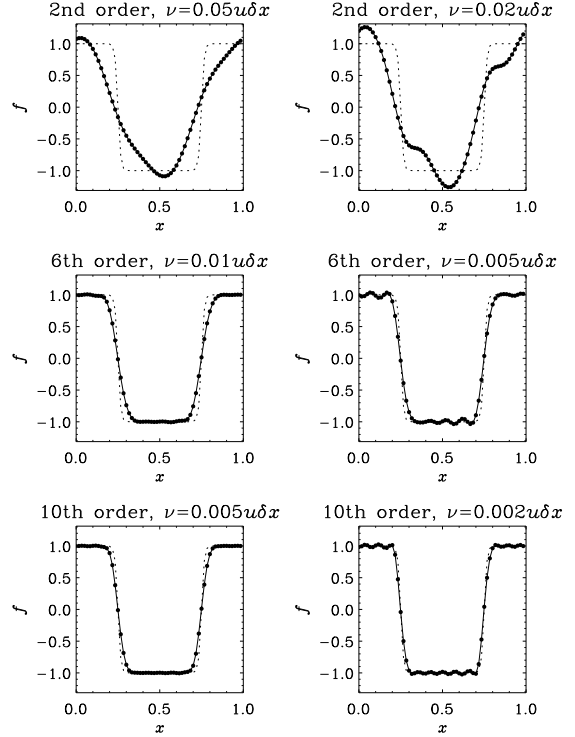}\caption{
Advection tests with schemes of different spatial order.
Resulting profile after advecting a step-like function 5 times
through the periodic mesh. The dots on the solid line give the location
of the function values at the computed meshpoints and the dotted line
gives the original profile. For the panels on the right hand side the
diffusion coefficient is too small and the profile shows noticeable
wiggles. $\delta x=1/60$.
}\label{Fpadvect_all_ccp2001}\end{figure}

\begin{figure}[t!]\centering\includegraphics[width=0.99\columnwidth]{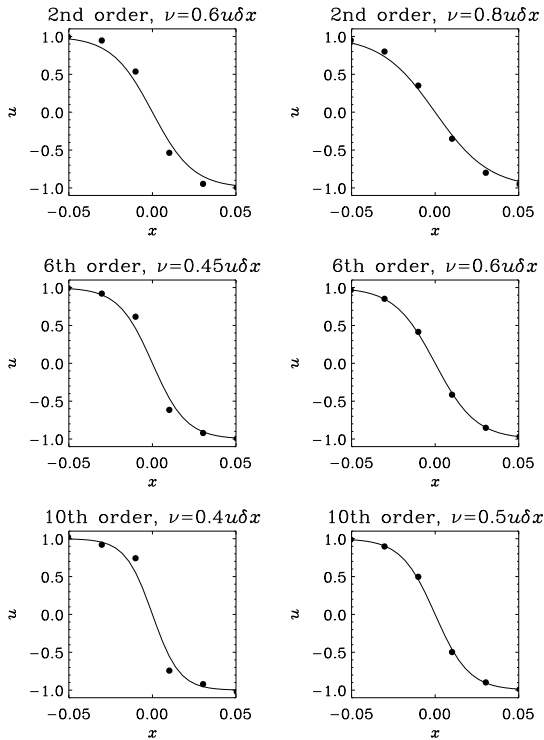}\caption{
Burgers shock with schemes of different spatial order and different
values of the viscosity. The solid lines give the analytic solution.
A second order scheme (top row) requires a
viscosity of at least $0.8u\delta x$, where $u$ is the amplitude of
the shock and $\delta x$ the mesh spacing. For a sixth-order scheme,
a viscosity of $0.6u\delta x$ yields good results, and for a tenth-order
scheme a viscosity of $0.5u\delta x$ can be used.
}\label{Fpburg}\end{figure}

It is commonly believed that high-order schemes lead to
Gibbs phenomena and that more viscosity is needed to
damp them out. In fact, the opposite is true as is demonstrated
in \Fig{Fpadvect_all_ccp2001} for advection tests and in
\Fig{Fpburg} for the stationary Burgers shock.

For the time stepping high-order schemes are necessary in order to reduce
the amplitude error of the scheme and to allow longer time steps. Usually
such schemes require large amounts of memory. However, there are the
memory-effective $2N$-schemes that require only two sets of variables
to be held in memory. Such schemes work for
arbitrarily high order, although not all Runge-Kutta schemes can be
written as $2N$-schemes \cite{2Nstorage,SH88}.
These schemes work iteratively according to the formula
\EQ
w_i=\alpha_i w_{i-1}+\delta t\,F(t_{i-1},u_{i-1}),
\EN
\EQ
u_i=u_{i-1}+\beta_i w_i.
\label{iterform0}
\EN
For a three-step scheme we have $i=1,...,3$.
In order to advance the variable $u$ from $u^{(n)}$ at time $t^{(n)}$
to $u^{(n+1)}$ at time $t^{(n+1)}=t^{(n)}+\delta h$ we set in \Eq{iterform0}
\EQ
u_0=u^{(n)}\quad\mbox{and}\quad u^{(n+1)}=u_3,
\EN
with $u_1$ and $u_2$ being intermediate steps. In order to be able to
calculate the first step, $i=1$, for which no $w_{i-1}\equiv w_0$ exists,
we have to require $\alpha_1=0$. Thus, we are left with 5 unknowns,
$\alpha_2$, $\alpha_3$, $\beta_1$, $\beta_2$, and $\beta_3$. Three
conditions follow from the fact that the scheme be third order, so we
have to have two more conditions. One possibility is the choose the
fractional times at which the right hand side is evaluated, for
example (0,~1/3,~2/3) or even (0,~1/2,~1). In the latter case the right hand
side is evaluated twice at the same time. It is therefore some sort of
predictor-corrector scheme. Yet another possibility is to require that
inhomogeneous equations of the form $\dot{u}=t^n$ with $n=1$ and 2 are
solved exactly.
The corresponding coefficients are listed in \Tab{Ttab_2N-RK3} and compared
with those given by Williamson \cite{2Nstorage}. In practice all of them
are about equally good when it comes to real applications, although
we found the first one in \Tab{Ttab_2N-RK3} (`symmetric') marginally better in some
simple test problems where an analytic solution was known.

\begin{table}[htb]\caption{
Possible coefficients for different $2N$-RK3 schemes.
}\vspace{12pt}\centerline{\begin{tabular}{lccccccc}
\hline
label & $\alpha_2$ & $\alpha_3$ & $\beta_1$ & $\beta_2$ & $\beta_3$ \\
\hline
symmetric           &  $-2/3$  &   $-1$   & 1/3 &  1  & 1/2 \\
predictor/corrector &  $-1/4$  &  $-4/3$  & 1/2 & 2/3 & 1/2 \\
inhomogeneous       & $-17/32$ & $-32/27$ & 1/4 & 8/9 & 3/4 \\
Williamson (1980)   &  $-5/9$  &$-153/128$& 1/3 &15/16& 8/15\\
\label{Ttab_2N-RK3}\end{tabular}}\end{table}

\section{Implementing magnetic fields}

Implementing magnetic fields is
relatively straightforward. On the one hand, the magnetic field causes
a Lorentz force, $\JJ\times\BB$, where $\BB$ is the flux density,
$\JJ=\nab\times\BB/\mu_0$ is the current density, and $\mu_0$ is the
vacuum permeability. On the other hand, $\BB$ itself evolves according
to the Faraday equation,
\EQ
{\partial\BB\over\partial t}=-\nab\times\EE
\label{induction}
\EN
where the electric field $\EE$ can be expressed in terms of $\JJ$ using
Ohm's law in the laboratory frame, $\EE=-\uu\times\BB+\JJ/\sigma$, where
$\sigma=(\eta\mu_0)^{-1}$ is the electric conductivity and $\eta$ is the
magnetic diffusivity.

In addition we have to satisfy the condition $\nab\cdot\BB=0$. This is
most easily done by solving not for $\BB$, but instead for the magnetic
vector potential $\AAA$, where $\BB=\nab\times\AAA$. The evolution
of $\AAA$ is governed by the uncurled form of \Eq{induction},
\EQ
{\partial\AAA\over\partial t}=-\EE-\nab\phi
\EN
where $\phi$ is the electrostatic potential, which takes the role of an
integration constant which does not affect the evolution of $\BB$.
The choice $\phi=0$ is most advantageous on numerical grounds. (By
contrast, the Coulomb gauge $\nabla\cdot\AAA=0$, which is very popular
in analytic considerations, would obviously be of no advantages, since
one still has the problem of solving a the solenoidality condition.

Solving for $\AAA$ instead of $\BB$ has significant advantages,
even though this involves taking another
derivative. However, the total number of derivatives taken in the code is
essentially the same. In fact, when centered finite differences are
employed, Alfv\'en waves are better resolved when $\AAA$ is used,
because then the system of equations for one-dimensional Alfv\'en waves
in the presence of a uniform $B_{x0}$ field in a medium of constant
density $\rho_0$ reduces to
\EQ
\dot{u}_z=(\mu_0\rho_0)^{-1}B_{x0}A_y'',\quad
\dot{A}_y=B_{x0}u_z,
\EN
where a second derivative is taken only once (primes denote
$x$-derivatives). If, instead, one solves for the $B_z$ field, one has
\EQ
\dot{u}_z=(\mu_0\rho_0)^{-1}B_{x0}B_z',\quad
\dot{B}_z=B_{x0}u_z',
\EN
where a first derivative is applied twice, which is far less accurate
at small scales if a centered finite difference scheme is used. At
the Nyquist frequency, for example, the first derivative is zero and
applying an additional first derivative gives still zero. By contrast,
taking a second derivative once gives of course not zero. The use of
a staggered mesh would circumvent this difficulty. However, such an approach
introduces additional complications which hamper the ease with which
the code can be adapted to new problems.

Another advantage of using $\AAA$ is that it is straightforward to
evaluate the magnetic helicity, $\bra{\AAA\cdot\BB}$, which is a
particularly important quantity to monitor in connection with dynamo
and reconnection problems.

The main advantage of solving for $\AAA$ is of course that one does
not need to worry about the solenoidality of the $\BB$-field, even
though one may want to employ irregular meshes or complicated
boundary conditions.

\section{Cache-efficient coding}

Unlike the CRAY computers that dominated supercomputing in the 80ties and
early 90ties, all modern computers have a cache that constitutes a significant
bottleneck for many codes. This is the case if large three-dimensional
arrays are constantly used within each time step. The advantage of this
way of coding is clearly the conceptual simplicity of the code. A more
cache-efficient way of coding is to calculate an entire timestep (or
a corresponding substep in a three-stage $2N$ Runge-Kutta scheme) only
along a one-dimensional pencil of data within the box. On Linux and Irix
architectures, for example, this leads to a speed-up by 60\%. An additional
advantage is a drastic reduction in temporary storage that is needed for
auxiliary variables within each time step.

\section{Large scale fields from helical turbulence}

Many astrophysical flows are affected by rotation and gravitational
stratification. These two effects can make the flows helical. In the sun,
for example, the kinetic helicity of the flow is negative in the northern
hemisphere and positive in the southern. It has long been known that
this can lead to the production of large scale magnetic fields \cite{KR80}.

\begin{figure}[t!]\centering\includegraphics[width=0.99\columnwidth]{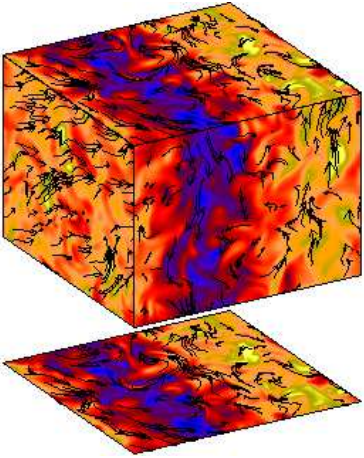}\caption{
Visualization of the magnetic field in a three-dimensional simulation
of helically forced turbulence. The turbulent magnetic field is
modulated by a slowly varying component that is force free.
}\label{FRun5_boxbot}\end{figure}

Recent simulations of helically forced turbulence have shown
that large scale fields are indeed produced \cite{Ref-2}. These large scale fields are
approximately force-free and of Beltrami type; see \Fig{FRun5_boxbot}.
The prototype of a Beltrami field is $\meanBB\propto(\cos z, \sin z, 0)$,
and it is easy to see that for this field $\meanJJ\times\meanBB=0$, where
$\meanJJ=\nab\times\meanBB/\mu_0$ is the current density.

\begin{figure}[t!]\centering\includegraphics[width=0.99\columnwidth]{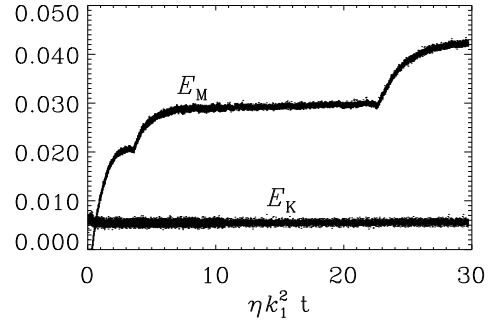}\caption{
Evolution of magnetic and kinetic energies, $E_{\rm M}$ and $E_{\rm K}$,
respectively, in a simulation with perfectly conducting boundaries.
}\label{Fpn}\end{figure}

At large scales, force-free Beltrami fields are
only possible in a periodic box, as considered in Ref.~\cite{Ref-2}.
When the box in non-periodic, the large scale field can still be
`nearly periodic'. In simulations with so-called `pseudo-vacuum' boundary
conditions, for example, a large scale field of the form
\EQ
\meanBB=(\cos\half z, \sin\half z, 0)\cos\half z
\EN
appeared; see Ref.~\cite{Ref-4}. We now consider the case of
perfectly conducting boundaries. Unlike the case of pseudo-vacuum
boundary conditions, where the energy of the large scale field
was somewhat below the kinetic energy, with perfectly conducting boundaries
the energy of the large scale field can strongly exceed the kinetic energy
of the turbulence; see \Fig{Fpn}.

The resulting super-equipartition of large scale magnetic energy is
primarily a consequence of the fact that the large scale field is
force-free and does not act back on the flow. Force-free fields
are generally helical and have to obey the equation of magnetic
helicity conservation,
\EQ
\dd\bra{\AAA\cdot\BB}/\dd t=-2\eta\mu_0\bra{\JJ\cdot\BB}.
\EN
The precise saturation level is obtained from the fact that in the steady
state the magnetic helicity, $\bra{\AAA\cdot\BB}$, is constant. This
implies that the current helicity, $\bra{\JJ\cdot\BB}$, vanishes.
Now, splitting the field into large and small scale components,
$\BB=\meanBB+\bb$, we have $\bra{\meanJJ\cdot\meanBB}=-\bra{\jj\cdot\bb}$.
Assuming that the fields are fully helical at small and large
scales, we have $\mu_0\bra{\meanJJ\cdot\meanBB}=\mp k_1\bra{\meanBB^2}$ and
$\mu_0\bra{\jj\cdot\bb}=\pm k_{\rm f}\bra{\bb^2}$, where $k_1$ is the
smallest wavenumber in the box and $k_{\rm f}$ is the wavenumber of
the forcing. (Upper/lower signs apply to positive/negative sign
of the helicity of the forcing function.) Due to non-periodic boundary
condition, however, the large scale field cannot be completely force-free.
Therefore we allow for an efficiency factor $\epsilon_{\rm LS}<1$ for
the large scales (LS). Thus, the final balance equation is
\EQ
\epsilon_{\rm LS}k_1\bra{\meanBB^2}=k_{\rm f}\bra{\bb^2},
\EN
i.e.\ the energy of the large scale field exceeds that of the small scale
field by a factor $\epsilon_{\rm LS}^{-1}k_{\rm f}/k_1$. This is consistent
with \Fig{Fpn}. Note, for comparison, that for fully periodic boxes,
$\epsilon_{\rm LS}=1$ and therefore the factor of superequipartion is only
$k_{\rm f}/k_1$.

The reason why the magnetic energy saturates in stages is connected with the
occurrence of different patterns of the large scale field: at early times the
large scale field is dominated by a pattern with a relatively large wavenumber
in one of the two horizontal directions (the boundaries are in the vertical
direction). Initially the large scale fields has 8 nodal planes, but then it
reduces to 4 and finally to 2 nodal planes.

Finally, it should be emphasized that we have discussed here only the
details of the saturation behavior. However, when the field is weak the
magnetic field grows always exponentially on a dynamical time scale
(usually over many orders of magnitude), and is independent of
magnetic helicity conservation.

\section{Conclusions}

The use of high-order schemes proved to be a useful compromise between
the cheap, but less accurate low-order methods and the
computationally more expensive spectral schemes. Explicit meshpoint
schemes can readily be implemented on massively parallel architectures
using High Performance Fortran (HPF) or the Message Passing Interface
(MPI). The $2N$-schemes of Williamson \cite{2Nstorage} are ideal for reducing
the amount of storage while still allowing the temporal order of the
scheme to be high.

\small
\vfill\bigskip\noindent{\it
$ $Id: paper.tex,v 1.4 2001/09/17 16:34:26 brandenb Exp $ $}

\end{document}